\def\eps{\varepsilon}
\def\vphi{\varphi}

\def\vk{\vec{k}}
\def\vpsi{\vec{\psi}}
\def\tr{\tilde{r}}
\def\tu{\tilde{u}}
\def\ds{\displaystyle}
\def\be{\begin{equation}}
\def\ee{\end{equation}}
\def\bea{\begin{eqnarray}}
\def\eea{\end{eqnarray}}

\documentclass[showpacs,preprint,aps,amsmath,eqsecnum]{revtex4}
\usepackage{epsfig}
\usepackage{graphics}

\begin{document}

\title{Unified static renormalization-group treatment of 
finite-temperature crossovers close to a quantum critical point}
\author{M.~T. Mercaldo, L. De~Cesare, I. Rabuffo}
\affiliation{Dipartimento di Fisica ``E.R. Caianiello",
Universit\`a di Salerno and CNISM , Unit\`a di
Salerno, I-84081 Baronissi (Salerno) Italy}
\author{A. Caramico~D'Auria}
\affiliation{Dipartimento di Scienze Fisiche,Universit\`a
di Napoli Federico II and  ``Coherentia", CNR-INFM, I-80125 Napoli, Italy}

\begin{abstract}
A nonconventional renormalization-group (RG) treatment close to
and below four dimensions is used to explore, in a unified and
systematic way, the low-temperature properties of a wide class of
systems in the influence domain of their quantum critical point.
The approach consists in a preliminary averaging over quantum
degrees of freedom and a successive employment of the Wilsonian RG
transformation to treat the resulting effective classical
Ginzburg-Landau free energy functional. This allows us to perform
a detailed study of  criticality of the quantum systems under
study. The emergent physics agrees, in many aspects, with the
known quantum critical scenario. However a richer structure of the
phase diagram appears with additional crossovers which are not captured by
 the traditional RG studies. In addition, in spite of the
intrinsically static nature of our theory, predictions about the
dynamical critical exponent, which parametrizes the link between
statics and dynamics close to a continuous phase transition, are
consistently derived from our static results.
\end{abstract}

\pacs{64.60.Ak.;05.70.Jk}

\maketitle

\section{Introduction}

Continuous quantum phase transitions (QPTs) constitute a very
topical subject in condensed matter physics
\cite{sachdev,qpts,sondhi,DCqpt,vojta-rev,GSI} and in the last few years their
intensive study  has stimulated  interesting speculations in other
branches of modern physics, too \cite{modernphys}. The fundamental
feature is that anomalous behaviors appear when a quantum critical
point (QCP) is approached. In particular, it is now well
established that the presence of such zero-temperature critical
points is the key to explain unsolved puzzles in the
low-temperature properties of many materials
\cite{heavyfermion,rutenati,HTSC}.

In spite of the large variety of systems that exhibit QPTs, their
critical properties can be described, following the seminal paper
by Hertz \cite{hertz} for itinerant magnets, using suitable
quantum Ginzburg-Landau (QGL) free energy functionals,
characterized by the dependence of the $n$-vector order parameter
field on the Matsubara-time variable $\tau$ and by the presence in
the free propagator of a term related to the intrinsic dynamic of
the original microscopic systems. This assures the correct
inclusion of the quantum degrees of freedom, by avoiding the
difficulties connected with noncommuting operators.

In the earliest works on the argument
\cite{hertz,young75,gerberbeck,lawrie,busDC80,gold79,noi-largeN,micnas} 
the effects of the zero-point
critical fluctuations  were essentially studied only at
temperature $T=0$ by applying the ideas of  Wilson's
renormalization-group (RG) approach.

The $T=0$ analysis does not produce additional conceptual
difficulties with respect to thermal phase transitions since
quantum  criticality is determined by the divergence of the length
scale, set by the correlation length $\xi$, as well as by the
divergence of the time scale $\tau_\xi\sim \xi^z$, where $z$ is
the dynamical critical exponent. The main conclusion is that, in
general, a QPT in $d$ dimensions is related to a classical
transition in $(d+z)$, except for Bose-like systems 
with ($-i\omega_l$) intrinsic dynamics (here $\omega_l$ denotes the usual 
bosonic Matsubara frequency: see next section). Relevant examples are the 
dilute Bose gas, the XY model in a transverse field and other models in the 
same quantum universality class \cite{sachdev,DCqpt,noi-largeN,noi97}. For these peculiar systems, 
an unusual ($T=0$) mean-field-like quantum criticality was found for $d<2$
 by variation of an appropriate control parameter (chemical potential, 
transverse magnetic field, and so on). This finding cannot be explained in 
terms of a simple dimensional crossover $d\to d+2$  but rather by means of 
a more complex crossover process $(d, n)\to (d= d+2,n=-2)$ involving also an
effective change of the dimensionality $n$ of the order parameter field: 
the quantum critical exponents of the Bose-like systems can be formally 
obtained from those for a classical $n$-vector model with dimensionality $d+2$
 and symmetry index $n = -2$ \cite{nota1}. This was conjectured in Ref.~\cite{busDC80} on 
the grounds of a ($T=0$) RG treatment up to second order in the natural 
expansion parameter $\eps=2-d$, proved to be valid to arbitrary order in 
$\eps$ for the interacting Bose gas \cite{uzu} and for the XY model in a transverse 
field \cite{kopec-chubu} and confirmed by exact large-$n$-limit calculations \cite{noi-largeN,DC82}.

It is also worth pointing out that for some systems of itinerant electrons, the  Hertz
theory \cite{hertz} does not seem properly adequate to describe
the correct $T=0$ critical behavior as will be specified in the
next section.  Anyway, in this paper we   consider  only systems
for which the QGL free energy functionals are expected to describe
correctly the quantum critical behavior.

The reliable and complete  description of finite-temperature
crossovers close to a QCP has a long history 
\cite{hertz,baba,schmeltz,walasek,crossovers,millis93,sachdevETAL,noi97,noi03,noi05,70s,DC82,walasek86,tvojta,sachdev97}. 
Wilsonian and field-theoretic RG treatments 
\cite{baba,schmeltz,walasek,crossovers,millis93,sachdevETAL,noi97,noi03,noi05}
 have been extensively used in
combination with nonperturbative and self-consistent methods 
\cite{sachdev,70s,DC82,walasek86,tvojta,sachdev97}.

The common opinion for a long period was that, at any finite
temperature, classical fluctuations control the behavior of the
system, but  it has become increasingly clear that the presence of
a QCP peculiarly influences   measurable quantities over a wide
range of  the low-temperature phase diagram. Indeed
 intricate crossovers between finite temperature
regimes may occur,  especially when there is a line of finite
temperature phase transitions ending in a QCP. Although  previous
partial RG investigations of low-temperature properties and crossovers
exist \cite{baba,crossovers}, the first detailed study of the low-temperature phase
diagram  around a QPT was performed by Millis
\cite{millis93} within a RG framework treating the thermal and
quantum fluctuations  on the same footing. He considered quantum
actions for itinerant antiferromagnets and ferromagnets and
depicted the corresponding low-temperature phase diagram by
solving the RG equations, step by step, in different regions
selected  by suitable conditions. The resulting description is
correct, but the derivation of the incoming crossover lines
appears cumbersome and, sometimes, rather unnatural.

In this paper we propose a nonconventional, but intrinsically
simple, approach to obtain a general and systematic  description
of the complex structure of the phase diagram when a QCP is
present, avoiding  the {\em step-by-step} Millis procedure.
Starting with a general QGL functional, (i) we integrate out the
degrees of freedom with nonzero Matsubara frequencies, thus
reducing the original quantum action to an effective classical one
with temperature-dependent coupling parameters, and then (ii) we
solve the related  RG equations to obtain the phase diagram and
the crossover scenario. It is worth mentioning that the first step
has been already used  by Sachdev \cite{sachdev,sachdev97} to
formulate a theoretical approach to finite-temperature quantum
criticality, which mixes perturbative predictions and known
$(T=0)$ RG results close to and above the quantum upper critical
dimension. In our picture, the temperature-dependent effective
couplings play a crucial role and we  show that, with this basic
ingredient, both the classical and quantum criticalities appear as
a natural result of the fusion of the classical world with the
underlying quantum one. This special feature, together with the
powerful Wilson RG method, allows us to draw out in a unified way
a series of low-temperature crossover lines, which separate
different asymptotic regimes, including some which do not
emerge in former approaches and which could be observable in
appropriate ultralow-temperature experiments.

Moreover, within our method, we avoid the direct control of the
quantum degrees of freedom in the various levels of approximation,
obtaining, at the end, the quantum criticality as an emergent
phenomenon. Just for this reason, we believe that the idea
developed in this paper may be conveniently employed in other
branches of theoretical physics, from quantum gravity to
cosmology.

The paper is structured  as follows. In Sec.~II we introduce the
quantum action which allows us to properly describe the
low-temperature critical properties of many systems exhibiting a
QCP. Then, after averaging over degrees of freedom with nonzero
Matsubara frequencies, we present the explicit expression of the
arising effective classical functional to one-loop approximation.
As a second step of our program, in Sec.~III the one-loop RG
equations for the temperature-dependent effective coupling
parameters are solved exactly close to and below four dimensions,
and the general expression of the correlation function as a
function of the temperature and of the original ``microscopic"
parameters is obtained for the quantum systems under study. Section~IV
is devoted to determine the critical-line equation in the
low-temperature regime and the related shift exponent. The
critical properties and the crossovers approaching the critical
line close to the $(T=0)$ ending point (here identified as a QCP)
are studied in Sec.~V and their unified description in terms of
two-parameters effective exponents is presented  in Sec.~ VI. In
the next Sec.~VII we localize other crossover lines far from
the phase boundary  to have a global picture of the phase diagram
for different quantum systems. Finally, in Sec.~VIII, some
conclusions are drawn.

\section{Quantum models and effective classical Hamiltonian}

A remarkable feature is that the critical properties of a wide
 variety of systems that exhibit QPTs can be described through a reduced number of 
QGL actions; each of them is representative of a given quantum
universality class, defined by the space dimensionality $d$, the
order parameter symmetry index $n$, and the dynamical critical
exponent $z$ that characterizes the intrinsic dynamics of the
systems in the class.

Bearing this in mind, in order  to be as  general as possible,  we
consider a quantum action which, in the Fourier space, is written
in the form
\bea \label{S1}
S\{\vpsi\}&=&S_0\{\vpsi\}+S_I\{\vpsi\}\\
S_0\{\vpsi\}&=&\frac12 \sum_{j=1}^n\sum_{\vk,\omega_l}
(r_0+k^2+\vphi(\vk,\omega_l))|\psi^j(\vk,\omega_l)|^2 \\
\label{S3} S_I\{\vpsi\}&=&\frac{u_0 T}{4 V} \sum_{i,j=1}^n
\sum_{\{k_\nu,\omega_{l\nu}\}} \! \! \! \! \delta_{\sum_{\nu=1}^4\vk_\nu;0}
\delta_{\sum_{\nu=1}^4\omega_{l\nu};0}
\psi^i(\vk_1,\omega_{l_1})\psi^i(\vk_2,\omega_{l_2})
\psi^j(\vk_3,\omega_{l_3})\psi^j(\vk_4,\omega_{l_4}). 
\eea 
Here
$\vpsi (\vk,\omega_l)\equiv \{\psi^j(\vk,\omega_l); j=1,...,n\}$
are the Fourier components of an $n$-vector real order parameter
field, $\vec k$ denotes a wave vector with a cutoff $\Lambda=1$
 (in convenient units), $T$ is the temperature, $V$ is
the volume,  and $\omega_l=2\pi l T$  $(l=0,\pm 1,\pm 2,...)$ are
the bosonic Matsubara frequencies. Of course, models with a
complex ordering field can be described in terms of $n=2m$ real
components (with $m=1,2,...$). The meaning of the coupling
parameters $r_0,u_0$ and the explicit expression of the function
$\vphi (\vec k, \omega_l) $, which defines the intrinsic dynamics,
depend on the physical system of interest. Our analysis can be
performed formally for a general $\vphi(\vk,\omega_l)$ and only
when it is necessary one can introduce its explicit expression to
have information about a particular quantum system. However, to be
specific, through this paper we focus on three basic models which
have recently attracted a great deal of attention to describe the
behavior close to a QCP of many materials subject to extensive
experimental studies in the latest years. They are characterized
by \cite{sachdev,qpts,sondhi,DCqpt,vojta-rev,GSI,hertz,young75,gerberbeck,lawrie,busDC80,gold79,noi-largeN,micnas,nota1,baba,schmeltz,walasek,crossovers,millis93,sachdevETAL,noi97,noi03,noi05,70s,DC82,walasek86,tvojta,sachdev97,rudne,Chakra,Kawa}:

(i) $\vphi(\vk,\omega_l)= \omega_l^2$. This function defines the
intrinsic dynamics of  the so-called transverse Ising-like systems
$(n\ge 1)$ \cite{Chakra} and allows one  to properly  describe, for
instance, the low-temperature properties of several magnetic
materials and compounds with quantum structural phase transitions
\cite{noi97}, for which the non-thermal control parameter is
related to the applied magnetic field and  the pressure;

(ii) $\vphi(\vk,\omega_l)= -i \omega_l$. This  is peculiar of the
class of Bose-like systems \cite{sachdev,DCqpt,noi-largeN,noi97} such as, for istance, 
those described by the
transverse XY model and antiferromagnetic dimer or ladder spin
materials where the field induced QPT  can be explained in terms
of a Bose-Einstein condensation  of magnons  \cite{Kawa};

(iii) $\vphi(\vk,\omega_l)= |\omega_l|$. This enters the action
model generally used for itinerant antiferromagnets \cite{sachdev,hertz,millis93}  
and other systems in the same quantum universality
class \cite{noi97}. In this context, it has been recently
speculated \cite{chubukov04} that the Hertz-Millis
 $\psi^4$-theory  \cite{hertz,millis93} of quantum criticality for itinerant
antiferromagnets is incomplete as it misses anomalous nonlocal
contribution to the interaction vertices (the effective bosonic
action  becomes nonlocal). Hence, it should fail to predict
results for dimensionalities $d\le 2$ with $z=2$. In contrast, for
$d > 2$ all the interaction terms are irrelevant  and the
Gaussian-like results preserve their validity. In any case, other
systems exist \cite{sachdev,noi-largeN,noi97} for which the
$\psi^4$- action (\ref{S1})-(\ref{S3}) with an
$|\omega_l|$-dynamics appears adequate.

A relevant feature is that our picture is quite general and may be
simply applied also to other quantum systems \cite{noi97} with
intrinsic dynamics described by $\vphi(\vec
k,\omega_l)=|\omega_l|^\mu/k^{\mu '} (\mu \ge 1,\mu ' \ge 0 $) in
the quantum action and with a dynamical critical exponent
$z=(2+\mu ')/ \mu$. The only concern is to calculate the Matsubara
frequency sums
$T\sum_{\omega_l}\left(r_0+k^2+\frac{|\omega_l|^\mu}{k^{\mu'}}\right)^{-1}$
which have been studied in Ref. \cite{noi97}, where examples of
other physical systems can be found. However, some caution must be
used for the relevant case $|\omega_l|/k$ usually assumed in the
action (\ref{S1})-(\ref{S3}) to describe quantum criticality of
clean itinerant ferromagnets \cite{sachdev,hertz,millis93}. It
has been indeed observed \cite{soft-modes} that the conventional
Hertz-Millis analysis  may predict incorrect results for $d\le 3$,
due to the existence of soft-modes at zero temperature that couple to the order parameter
field and thus preclude the  construction of a conventional QGL
action. A more recent study based on a entirely different  point
of view \cite{Chubu} seems indeed to confirm the non-validity of
the Hertz-Millis theory for $(d\le 3)$ dimensional itinerant
ferromagnets.

 Now, we have all the basic ingredients to start with our proposal.

Our first step is to average over the degrees of freedom
with $\omega_l\neq 0$ to generate an effective classical
functional, where the quantum nature of the original action enters
 the new temperature-dependent coupling parameters as a result of the
averaging process.

For this purpose we separate in the free action the term with
$\omega_l=0$, writing
 \be
S\{\vpsi\}=S_0\{\vec\Phi\}+S_0\{\vpsi(\vk,\omega_l\neq0)\}+S_I\{\vec{\Phi};\vpsi(\vk,\omega_l\neq0)\}\;,
\ee
where $\vec\Phi(\vk)=\vpsi(\vk,\omega_l=0)$. Then the
partition function $Z=\int{\cal D}[\vpsi] e ^{-S\{\vpsi\}}$ can be
written as
 \bea
\label{Z1} Z&=& \int{\cal D}[\vec\Phi]\left\{e^{-S_0\{\vec\Phi\}}
\int {\cal D}[\{\vpsi(\vk,\omega_l\neq0)\}]
e^{-\left[S_0\{\vpsi(\vk,\omega_l\neq0)\}+S_I\{\vec\Phi;\vpsi(\vk,\omega_l\neq0)\}\right]}\right\}\\
\nonumber
 &\equiv& \int{\cal D}[\vec\Phi]e^{-{\cal H}\{\vec\Phi\}}\;,
\eea
 where ${\cal H}\{\vec\Phi\}$ denotes the dimensionless effective classical
Hamiltonian which arises from the reduction procedure of the
quantum degrees of freedom. Working within a perturbative scheme
to one-loop approximation and with the condition  ${\cal
H}\{\vec\Phi=0 \}\equiv 0$, we find for ${\cal H}\{\vec\Phi\}$ the
$\Phi ^4$-expression
\bea
\label{H1} {\cal H}\{\vec\Phi\}&=&\frac{1}{2}
\sum_{j=1}^n\sum_{\vk} (\tilde r_0+k^2) |\Phi^j(\vk)|^2\\
\nonumber &+& \frac{\tilde u_0}{4V} \sum_{i,j=1}^n
\sum_{\{\vk_\nu\}} \delta_{\sum_{\nu=1}^4 \vk_\nu;0}
\Phi^i(\vk_1)\Phi^i(\vk_2)\Phi^j(\vk_3)\Phi^j(\vk_4)\;. \eea

\begin{figure}[b]
\vspace{0.5cm}
  \begin{center}
    \includegraphics[width=7cm]{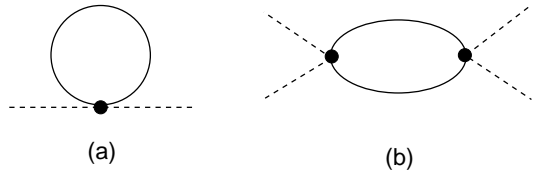}
  \end{center}
  \caption{Diagrams leading: (a) to the effective coupling $\tr_0$;
 and (b) to the new quartic coupling $\tu_0$. Here the dashed lines
correspond to $\vpsi(\vk,0)$, the straight lines to
$\vpsi(\vk,\omega_l\neq0)$.}
  \label{Fig1}
\end{figure}

The diagrams that contribute to the effective coupling parameters
$\tilde r_0$ and $\tilde u_0$ in Eq.~(\ref{H1}) are shown in Fig.~1.
These parameters are connected to the original microscopic ones
$ r_0$ and $ u_0$ by the  relations (with $(1/V)\sum_{\vk}(...)
  \longrightarrow K_d\int_0^1 dk k^{d-1}(...)$ and
  $K_d=2^{1-d}\pi^{-d/2}/\Gamma(d/2)$)
\bea
\label{in.cond.1}
\tilde r_0&=&r_0+K_d(n+2)u_0\int_0^1dk k^{d-1} \left\{T
 \sum_{\omega_l\neq0} G_0(\vk,\omega_l)\right\}\;, \\
\label{in.cond.2}
\tilde u_0&=&T u_0 \left [1-K_d{{n+8}\over 2}
u_0 \int_0^1 dk k^{d-1}\left\{T \sum_{\omega_l\neq 0}
 G_0(\vk,\omega_l) G_0(-\vk,-\omega_l)\right\} \right ],
\eea
where
 \be
\label{prop} G_0(\vk,\omega_l)=\frac1{r_0+k^2+\vphi(\vk,\omega_l)}
\ee is the free propagator which takes memory of the distinctive
features of a quantum system.

Defining
 \be
\label{GT} G(r_0,T)=\int_0^1 dk k^{d-1} \left \{ T \sum_{\omega_l}
G_0(\vk,\omega_l)\right \}\;,
\ee
Eq.~(\ref{in.cond.1}) can be
conveniently written as
\be
\label{errezerotilde}
\tr_0=r_0+K_d(n+2)u_0 \left \{G(r_0,T)- T
\int_0^1 dk \frac{k^{d-1}}{r_0+k^2}\right \} \;,
\ee
where the last
term on the right hand side (r.h.s) represents the contribution of
the zero-frequency term.

Eqs.~(\ref{H1})-(\ref{errezerotilde}) will play a key  role for
the next developments.


\section{One-loop RG equations for the effective  classical Hamiltonian
and their solution close to and below four dimensions}

One can now  apply the standard RG approach to the effective
classical Hamiltonian (\ref{H1}), which represents the $n$-vector
model. The results are well known and to one-loop approximation,
where the Fisher correlation length exponent $\eta=0$, the
appropriate flow equations are
\bea
\label{RGeq1}
\frac{d \tr(l)}{dl}&=& 2 \tr(l) + K_d (n+2)  \frac{\tu(l)}{1+\tr(l)} \;,\\
\label{RGeq2} \frac{d \tu(l)}{dl}&=& (4-d)\tu(l) - K_d (n+8)
\frac{\tu^2(l)}{(1+\tr(l))^2} \;,
\eea
to be solved with the initial conditions:
\be
\tr(l=0)=\tr_0(T,r_0,u_0),\quad \tu(l=0)=\tu_0(T,r_0,u_0).
\ee

Of course the explicit dependence of the initial effective
parameters on the temperature and the physical parameters $r_0,
u_0$ reflects the microscopic nature of the quantum models here
considered. It is worth noting that, after reducing to the effective Hamiltonian,  
the physical temperature does not enter explicitly in the 
RG machinery. Hence it is not involved in the renormalization procedure, 
which acts only on the effective coupling
parameters, but appears at the end of calculations  
through the initial conditions to be
used for solving the RG recursion relations.

Without making explicit reference to the underlying fixed-point
scenario, we will adopt here the point of view that the RG
transformation is also a systematic step-by-step averaging
procedure to obtain (at a given level of approximation) the
partition function of a macroscopic system, which takes properly
into account the competing effects of classical and quantum
fluctuations. In our picture, the different microscopic dynamics
will emerge without involving the dynamical critical exponent $z$
as happens , in contrast, in the RG treatments based directly on
path-integral representations \cite{sachdev,qpts,sondhi,DCqpt,vojta-rev,GSI}. Bearing
this in mind, with the aim of exploring the low-temperature
properties of a quantum system, we need to solve
Eqs.~(\ref{RGeq1})-(\ref{RGeq2}) exactly to order of interest.

The solution for $\tilde u (l)$ to first order in  $\eps=4-d$ is
\be
\tu(l)= \frac{\tu_0 e^{\eps l}}{1+(n+8)K_d
(\tu_0/\eps)(e^{\eps l}-1)}\;.
\ee
This allows us to obtain in a simple form the
appropriate solution for the relevant coupling parameter $\tr(l)$
 through the combination \cite{rudne}

\be
t(l)=\tr(l)+\frac12 (n+2) K_d \tu(l)-\frac12 (n+2) K_d \tu(l)
\tr(l) \ln[1+\tr(l)]\;
\ee
which scales as
\be
\label{tl}
t(l)=e^{\Lambda(l)} t(l=0),
\ee
where
\be
\label{tl=0}t(l=0)= \tr_0+\frac12(n+2)K_d \tu_0 -\frac12 (n+2)
K_d \tu_0 \tr_0 \ln(1+\tr_0)
\ee
and
\be
\label{Lambdal}
\Lambda(l)=2l - \frac{n+2}{n+8} \ln[1+(n+8) K_d (\tu_0/\eps)(e^{\eps l} -1)]\;.
\ee

Hereafter we consider the most interesting case $\eps>0$ $(d<4)$.

As a conclusion of this section, we write down the initial expression
$t(l=0)$ of the non-linear scaling field (\ref{tl}) in a low-temperature
form very convenient for the next developments.

Working to first order in the original coupling parameters $r_0, u_0$,
Eq.~(\ref{tl=0}) yields
\be \label{tl0} t(l=0) = r_0 + K_d (n+2) u_0 G(r_0,T) \equiv
t(r_0,T)\;,
 \ee
 where, with  the notation $t(r_0,T)$,  we have
explicitly introduced the dependence on the temperature $T$ and on
the microscopic non-thermal parameter $r_0$.

Moreover, performing the sums over Matsubara frequencies in Eq.~(\ref{GT}), 
in the low-temperature limit  $G(r_0,T)$ can be
written as
\be 
\label{Gr0t} 
G(r_0,T)\simeq G(r_0,0)+ T^\psi {\cal G}(r_0/T^{2/\zeta})\;. 
\ee 
The value of the  exponents $\psi$ and
$\zeta$, whose physical meaning will become clear later, together
with  the explicit expressions of $ G(r_0,T)$ and ${\cal G} (r_0
/T^{2/\zeta})$ for the different quantum models considered in this
paper, are collected in Table~\ref{T1}.

\begin{turnpage}
\begin{table}
\caption{Values of the exponents $\psi$ and $\zeta$ and explicit expressions of the functions
$G(r_0,T)$ and ${\cal G}(r_0,T^{2/\zeta})$ for the three classes of systems we are investigating.}
\label{T1}
\begin{tabular}{| c | c | c | l | l |}
\hline
$\vphi(\vk,\omega_l)$ &  $\psi$ & $\zeta=\frac{d-2}{\psi-1}$ & $\qquad \qquad G(r_0,T)$ & $\qquad \qquad{\cal G}(r_0/T^{2/\zeta})$ \\
\hline
$\omega_l^2$ &  $d-1$ & 1 & $K_d\ds\int_0^1 k^{d-1}\frac12
\frac{{\rm cth}[(1/2T)\sqrt{r_0+k^2}]}{\sqrt{r_0+k^2}}$ & $ \ds\int_0^\infty dx \frac{x^{d-1}}{[(r_0/T^2)+x^2]^{1/2}}
\frac1{\exp[(r_0/T^2)+x^2]^{1/2}-1}$\\ & & & & \\
$-i \omega_l$ &  $\ds\frac d2$ & 2 & $K_d\ds\int_0^1 k^{d-1}\frac12
{\rm cth}\frac{r_0+k^2}{2T}$ & $\ds\frac12 \ds\int_0^{\infty} dx \frac{x^{d/2-1}}{\exp[(r_0/T)+x]-1}$\\  & & & & \\
$|\omega_l|$ &  $\ds\frac d2$ & 2  & $K_d\ds\int_0^1 k^{d-1}
\int_0^1 \frac{d\omega}\pi \left[{\rm cth}\frac{\omega}{2T}\right]
\frac\omega{(r_0+k^2)^2+\omega^2}$ & $\ds\int_0^\infty \frac{dy}\pi
\int_0^1 dx \frac x{e^x-1}\frac{y^{d/2-1}}{[(r_0/T)+y]^2+x^2}$\\
\hline
\end{tabular}
\vspace{0.6cm}
\end{table}
\end{turnpage}

Finally we write the low-temperature expression for $t(r_0,T)$ 
\be
\label{tr0T-prima}
t(r_0,T)\simeq r_0 + K_d(n+2) u_0 [G(r_0,0)+
T^\psi {\cal G}(r_0/T^{2/\zeta})]\;,
\ee
which will play a
relevant role in the next analysis.

\section{Critical line}

From the rescaling relation (\ref{tl}) for the relevant field
$t(l)$, one immediately has that the critical line in the
$(r_0,T)$-plane, close to and below four dimensions, is determined
by the condition $t(r_0,T)=0$ which yields the critical line
equation
\be
\label{lineacritica}
r_0+K_d(n+2)u_0G(r_0,T)=0 \;.
\ee
Then, solving this equations with respect to $r_0$ or  $T$, to first order in the
coupling parameters, we obtain the following equivalent
low-temperature representations of the critical line
\bea
\label{r0c}
r_{0c}(T)&\simeq&r_{0c}-A(n,d)u_0 T^\psi\;, \\
\label{Tc}
T_c(r_0)& \simeq&
[A(n,d)u_0]^{-1/\psi}(r_{0c}-r_0)^{1/\psi}\;, \quad (r_0\leq
r_{0c})\;,
\eea
where
\be \label{defs} r_{0c}=-K_d(n+2) u_0 G(0,0) \; ,\quad
A(n,d)=(n+2) K_d {\cal G}(0)\;, 
\ee 
\begin{table}[b]
\caption{Explicit expressions of the quantities that enter the critical line equation.}
\label{T2}
\begin{center}
\begin{tabular}{| c | c | c |}
\hline
$\vphi(\vk,\omega_l)$ & $r_{0c}$ & $A(d,n)$\\
\hline
$\omega_l^2$ & $-\ds\frac{(n+2)K_d u_0}{2(d-1)}$ & $(n+2)K_d \Gamma(d-1)\zeta(d-1)$\\
$-i\omega_l$ & $-\ds\frac{(n+2)K_d u_0}{2d}$ & $\ds\frac{n+2}2 K_d \Gamma(d/2)\zeta(d/2)$\\
$|\omega_l|$ & $\ds\frac{n+2}{2\pi d}K_d u_0\left(\ln2-\beta(\frac{d+2}4)+\frac4d\right)$ &
 $\ds\frac{n+2}2 K_d \frac{\Gamma(d/2)\zeta(d/2)}{\sin(\frac{\pi d}4)}$\\
\hline
\end{tabular}
\vspace{0.6cm}
\end{center}
\end{table}
whose explicit expressions,
for the models here considered, are given in Table~\ref{T2}.
Eq.~(\ref{r0c}) (or (\ref{Tc})) shows that, for different quantum
models,   the critical line ends in the point ($r_0=r_{0c},T=0$)
that, as will be clear from the following analysis, plays  just
the role of a QCP. This feature makes clear the physical meaning
of the parameter $\psi$ as the phase boundary exponent, also known
as the shift exponent, with values reported in Table~\ref{T1}. In
particular, by extrapolation to $d=3$, one has $\psi =2$ for
$\omega_l^2$-intrinsic dynamics and $\psi = 3/2$ for cases
$(-i\omega_l,|\omega_l|)$.

It is worth noting that the
low-temperature shape of the phase boundary  is strictly related
to the microscopic nature of the system under study as a result of
the quantum degrees of freedom reduction procedure performed in
Sec. II.

For future convenience it is useful to  express the initial field
$t(r_0,T)$,
 in terms of the critical line equation $r_{0c}(T)$ or $T_c(r_0)$
as follows:
\be
\label{tr0T} t(r_0,T)=[r_0-r_{0c}(T)]+ K_d
(n+2)u_0[G(r_0,T)-G(r_{0c}(T),T)], 
\ee
or
\be
\label{tr0Tbis} t(r_0,T)= K_d
(n+2)u_0[G(r_0,T)-G(r_{0},T_c(r_0))] \;. 
\ee

\section{Low-temperature critical properties and crossovers}

In this section we study the low-temperature behavior of some
relevant quantities, e.g. the correlation length and the
susceptibility, when one approaches the critical line following
different thermodynamic paths in the phase diagram.

As usual in the RG approach, the correlation length $\xi$ and the
susceptibility $\chi$ can be expressed as \cite {rudne}
\begin{equation}
\xi=\xi_0e^{l^\ast}, \qquad  \chi=\chi_0e^{2l^\ast}
\end{equation}
where $l^{\ast}=l^{\ast}(r_0,T)\gg 1$ is determined by the
condition $t(l^\ast)=1$. By using Eqs.~(\ref{tl}), (\ref{Lambdal})
and (\ref{tl0}) we find for $l^\ast$, to order of interest in the
microscopic coupling parameters, the self-consistent equation
\begin{equation}
e^{2l^\ast}[1+(n+8)K_d (u_0/\eps)T(e^{\eps
l^\ast}-1)t(r_0,T)]^{-{{n+2}\over {n+8}}}=1 \;,
\end{equation}
which has the low-$T$ solution
\begin{equation}
\label{elstar} e^{l^\ast}=[t(r_0,T)]^{-{1\over 2}}\left
\{1+(n+8)K_d (u_0/\eps)T[t(r_0,T)]^{-{\eps\over 2}}\right
\}^{{n+2}\over {2(n+8) }} \;,
\end{equation}
yielding directly the dimensionless correlation length
$\xi/\xi_0$.

 Eq.~(\ref{elstar}) contains all the physics of interest for us.
It allows us to calculate not only the correlation
length and the susceptibility, but also the singular part of the
free energy density $F_s(r_0,T)$ through the usual scaling
relation $ F_s(r_0,T)\sim e^{-dl^\ast}\sim \xi^{-d}$ and hence the
singular part of the specific heat ${{C_s(r_0,T)/
T}}=-{{\partial^2 F_s(r_0,T)}/{\partial T^2}}$.

\subsection {Critical behavior close to the phase boundary and the
Ginzburg line.}

 Near the critical line within the disordered phase
($t(r_0,T)>0$), the last term in the r.h.s. of Eq.~(\ref{tr0T})
can be neglected to the order of interest in the parameters $r_0$
and $u_0$, and hence the field $t(r_0,T)$ in this region  assumes
the simplest form
\begin{equation}
\label{tr0T2} t(r_0,T)\simeq r_0-r_{0c}(T) \;,
\end{equation}
and measures, at any given temperature $T$, the horizontal
distance from the critical line.

At $T=0$ and $r_0\rightarrow r_{0c}^+ $,  from eqs.~(\ref{elstar})
and ~(\ref{tr0T2}) one has
\begin{equation}
\label{xi-chi} \xi\sim (r_0-r_{0c})^{-{1\over2}} \qquad , \;
\chi\sim (r_0-r_{0c})^{-1}\;.
\end{equation}
These $(T=0)$-mean field (MF) results, just expected for quantum
systems above their upper critical dimension $d_{cu}^{(q)}=4-z$ (i.e.
when $d+z>4$), allow us to interpret the point $(r_0=r_{0c},T=0)$
in the phase diagram as the QCP of the different quantum models
here considered (see Table~\ref{T2} for explicit values of
$r_{0c}$) where the correlation length and susceptibility diverge.

Notice that the result (\ref{xi-chi}) is appropriate for quantum
systems with $z>1$ also for $d=3$. An exception occurs for case
$z=1$ ($\omega_l^2$- dynamics) for which the quantum upper
critical dimension is  $d_{cu}^{(q)}=3$ and logarithmic
corrections to MF results are expected at $d=3$. In any case no
inconsistency enters the problem because our static theory is
really valid only close to and below four dimensions and caution
must be used when one extrapolates the results to $d=3$.

At finite temperature, Eq.~(\ref{elstar}) provides two different
asymptotical behaviors for $e^{l^\ast}$ according to which term is
dominant in the brackets. One has therefore for  the correlation
length $\xi$ and the susceptibility $\chi$, as $r_0\rightarrow
r^+_{0c}(T)$  along a thermodynamical path parallel to the $r_0$
axis, the asymptotical behaviors \cite{footnote}
\begin{equation}
\label{5.6}
\xi\sim(r_0-r_{0c}(T))^{-\nu_r} \qquad , \;
\chi\sim (r_0-r_{0c}(T))^{-\gamma_r}
\end{equation}
with
\be
\label{5.7}
 \nu_r=\left\{
\begin{array}{l}
 {1\over 2}\;, \qquad \quad\qquad\qquad\qquad  \;\, {\rm if} \quad
r_0\gg r_{0c}(T)+ [K_d(n+8) (u_0/\eps)T]^{2/\eps} \\
{1\over 2}\left(1+{n+2\over{2(n+8)}}\eps\right)\;, \quad
{\rm if} \quad r_{0c}(T) < r_0 \ll r_{0c}(T)+ [K_d(n+8)
(u_0/\eps)T]^{2/\eps}\;,
\end{array} \right.
\end{equation}
and $\gamma_r=2\nu_r$ since, in our one-loop analysis, the Fisher
exponent $\eta=0$. From now on we consider explicitly only the
critical exponents for the correlation length being $\chi \sim
\xi^2$ in any case.

Eq.~(\ref{5.7}) suggests that, varying the distance from the
critical line decreasing $r_0$ towards $r_{0c}(T)$ at fixed $T$,
the system undergoes a crossover from a MF behavior to a classical
Wilsonian (W) one, except at $T=0$ where a MF behavior is expected
as $r_0\rightarrow r_{0c}^+$ (see Eq.~(\ref{xi-chi})). The
crossover line  determined by
\begin{equation}
\label{rogi}
 r_{0Gi}(T)=r_{0c}(T)+[K_d (n+8)(u_0/\eps)T]^{2/\eps}, \;
\end{equation}
will be  called the ``Ginzburg line". It is worth noting that the
 horizontal
distance between the Ginzburg line and the critical one
\begin{equation}
t(r_{0Gi}(T), T)=r_{0Gi}(T)-r_{0c}(T)\;,
\end{equation}
goes to zero decreasing the temperature according to a power-law
with exponent $2/\eps$ (independent of the particular model) and
hence both the critical and Ginzburg lines merge at the QCP.

In terms
of $t(r_{0Gi}(T), T)$, Eq.~(\ref{elstar}) can be conveniently
written as
\begin{equation}
\label{5.10} e^{l^\ast}=[t(r_0,T)]^{-{1\over
2}}\left\{1+\left[{{t(r_{0Gi}(T),
T)}\over{t(r_0,T)}}\right]^{\eps/2}\right\}^{{n+2}\over {2(n+8)}}
\; .
\end{equation}
In order to evaluate the effective correlation length exponent
which interpolates between the two regimes in Eq.~(\ref{5.7}) and hence
to describe the previous crossover, it is natural to express
the correlation length  in terms of the renormalized distance from
the critical line at fixed $T$
\begin{equation}
\label{5.11}
 x={{t(r_0,T)}\over {t(r_{0Gi}(T), T)}}=
{{r_0-r_{0c}(T)}\over {r_{0Gi}(T)-r_{0c}(T)} }\;,
\end{equation}
as
\begin{equation}
\label{5.12} \xi=\xi_0\left[t(r_{0Gi}(T), T) \right]^{-{1\over
2}}h(x),
\end{equation}
where the scaling function $h(x)$ is given by
\begin{equation}
\label{5.13} 
h(x)=x^{-{1\over 2}}\left(1+x^{-{\eps\over
2}}\right)^{{n+2}\over {2(n+8)}}.
\end{equation}
Then,  from Eqs.~(\ref{5.10})-(\ref{5.13}) it is easy to obtain the required
effective exponent
\be
\label{5.14}
\nu_r^{\rm eff}(x)=- \frac{d \ln h(x)} {d \ln x}
= \frac{1}{2}\left
[1+\frac{n+2}{2(n+8)}\eps\left(\frac{1}{1+x^{\eps/2}}\right)\right] \;,
\ee
 which reproduces the asymptotic values in Eq.~(\ref{5.7})   in the
limiting cases $x\gg 1$ and $x\ll 1$, respectively.

As concerning the singular part of the free energy  density, we
immediately have for $r_0\rightarrow r_{0c}^+(T)$
\be
\label{5.15} F_s(r_0,T)\sim (r_0-r_{0c}(T))^{d\nu_r}  \;,
\ee
and hence for the specific heat we get
\begin{equation}
\label{5.16}
\frac{C_s(r_0,T)}{T}\sim
\left(r_0-r_{0c}(T)\right)^{-\alpha_r}  \;,
\end{equation}
with
\be
\label{5.17}
\alpha_r\simeq \left\{
\begin{array}{l}
 0 \;, \qquad \qquad \;\, {\rm if} \quad r_0 \gg
r_{0Gi}(T)\\
\frac {4-n}{2(n+8)}{\eps} \;, \qquad {\rm if} \quad r_{0c}(T)< r_0\ll
r_{0Gi}(T)  \;.
\end{array} \right.
\end{equation}

Of course, also in this case one can define and easily calculate
an effective specific heat exponent $ \alpha_r^{\rm eff}(x)$,
whose rather cumbersome expression is, however, inessential for
our purposes.

A similar analysis can be performed approaching the critical line
 along  thermodynamic paths parallel to the $T$-axis in the phase
 diagram.
For this purpose it is necessary to assume  for $t(r_0,T)$ the
representation (see Eq.~(\ref{tr0Tbis}))
\begin{equation}
\label{5.18}
t(r_0,T)\simeq A(n,d)u_0  \left(T^\psi-T^\psi_c(r_0)\right) \;.
\end{equation}
First, we suppose  $r_0\neq r_{0c}$ so that $T_c(r_0)\neq 0$ and
Eq.~(\ref{5.18}) reduces to
\begin{equation}
\label{5.19}
 t(r_0,T)\simeq\psi[T_c(r_0)]^{\psi-1}A(n,d)u_0(T-T_c(r_0)),
\end{equation}
which  measures, at any fixed $r_0<r_{0c}$, the vertical
distance from the critical line. Then, when $T\rightarrow
T_c^+(r_0)$, Eq.~(\ref{elstar}), together with (\ref{5.19}),
provides for the susceptibility and the correlation length the
asymptotic behaviors
\begin{equation}
\label{5.20} \xi\sim (T-T_c(r_0))^{-\nu_T} \qquad, \qquad \chi\sim
(T-T_c(r_0))^{-\gamma_T} \,,
\end{equation}
with
\be
 \label{5.21}
\nu_T \simeq
 \left\{
 \begin{array}{l}
 \frac12 \;,\qquad \qquad  \qquad\qquad {\rm if} \quad T \gg T_{Gi}(r_0)\\
\frac12\left[1+\frac{n+2}{2(n+8)}\eps\right]\;, \qquad {\rm if}
\quad T_c (r_0)<T \ll T_{Gi}(r_0)  \;,
\end{array} \right.
\end{equation}
 and $\gamma_T=2\nu_T$. Here
 \be
\label{5.22}
T_{Gi}(r_0)=T_c(r_0)+[\psi
A(n,d) u_0]^{-1}[(n+8)K_d (u_0/\eps)]^{2/\eps}
[T_c(r_0)]^{2/\eps}\;,
\ee
is the temperature-representation of the Ginzburg line.
The extrapolation of this result to $d=3$, according to  the 
genuine Wilson RG philosophy, yields a $T_c^2$ 
deviation from the critical line for any quantum system, 
consistently with the ($d=3$)-prediction of Ref.~\cite{millis93}.

 Note the
coincidence of the exponents $\nu_r$ and $\nu_T$ obtained along
horizontal and vertical thermodynamic  paths, respectively
\cite{footnote}, for $r_0<r_{0c}$. This is a consequence of the
linearization (\ref{5.19}) valid only when $T_c(r_0)$ is finite.

A suitable form for  the dimensionless correlation length
(\ref{elstar}) is now
\be
\label{elstar2}
e^{l^*}=[t(r_0,T)]^{-1/2}\left[1+\frac T{T_{Gi}(r_0)}
\left(\frac{t(r_0,T_{Gi}(r_0))}{t(r_0,T)}\right)^{\eps/2}\right]^{\frac{n+2}{2(n+8)} }\; ,
\ee
where
\bea
\label{5.24}
t(r_0,T_{Gi}(r_0))&=&\psi[T_c(r_0)]^{\psi-1} A(n,d) u_0 [T_{Gi}(r_0)-T_c(r_0)]\\
\nonumber &=&[(n+8)K_d (u_0/\eps)T_c(r_0)]^{2/\eps} \eea
estimates, at fixed $r_0<r_{0c}$, the vertical distance between
the Ginzburg line and the critical one. In terms of the crossover
parameter $x=[t(r_0,T)]/[t(r_0,T_{Gi}(r_0)]$, the correlation
length (\ref{elstar2}) looks like
\be \label{xi} \xi=\xi_0[t(r_0,T_{Gi}(r_0))]^{\frac12}h(x), \ee in
terms of the same scaling function (\ref{5.13}) that appears in
the representation (\ref{5.12}). So we recover, along a path
parallel to the $T$-axis, an effective exponent $\nu_T^{\rm
eff}(x)$ of the form (\ref{5.14}).

For the singular part of the free energy density we can now write
\be
F_s(r_0,T)\sim (T-T_c(r_0))^{d \, \nu_T}  \;. \ee
In this way
we obtain for the specific heat, along a thermodynamical path
parallel to the $T$-axis, the expression
\be
\frac{C_s(r_0,T)}T \sim (T-T_c(r_0))^{-\alpha_T} \;,
\ee
with asymptotic critical exponents
\be
\label{alfat}
\alpha_T\simeq \left\{
\begin{array}{l}
0 \;,\qquad \qquad \;\, {\rm if} \qquad  T \gg T_{Gi}(r_0)\\
\frac{4-n}{2(n+8)}\eps\; ,  \qquad {\rm if} \quad T\ll T_{Gi}(r_0) \;,
\end{array} \right.
\end{equation}
which are identical to the previous ones obtained for horizontal paths. Of
course, we  have also $\alpha_T^{\rm eff}(x)\equiv
\alpha_r^{\rm eff}(x)$.

\subsection{Critical behavior along the quantum critical trajectory $(r_0=r_{0c},T\to 0)$.}
\label{VB}

Experimental informations that characterize the low-temperature
behavior of a quantum system can be obtained fixing the
non-thermal control parameter $r_0$ at its QCP value $r_{0c}$ and
decreasing the temperature along the so-called \cite{qpts} quantum
critical  trajectory. Hence, this case deserve a particular
attention.

At $r_0=r_{0c}$ and $T\rightarrow 0$ the field (\ref{tr0T2})
becomes
\be
\label {tr0cT}
t(r_{0c},T) = A(n,d) u_0 T^\psi,
\ee
so that
the dimensionless correlation length (\ref{elstar}) takes the form
\be
\label{elstar3}
e^{l^*}=[A(n,d) u_0]^{-1/2} T^{-\psi/2}
\left[1+\left(\frac
T{T^*}\right)^\phi\right]^{\frac{n+2}{2(n+8)}},
\ee
where we have
defined the characteristic temperature $T^*$ and the exponent $\phi$ as
\bea
\label{Tstar}
 T^* &=& \left[\frac \eps{(n+8)K_d
u_0}\right]^{\frac 1\phi}
[A(n,d) u_0 ]^{\frac \eps{2 \phi}},\\
\label{fi}
 \phi &=& 1-\frac \eps 2 \psi \;.
\eea
Eq.~(\ref{elstar3})
provides two different asymptotical behaviors decreasing the
temperature towards the QCP. Defining indeed  for the correlation
lenght and susceptibility the critical exponents $\nu_T$ and
$\gamma_T$ as $\xi \sim T^{-\nu_T}, \quad \chi \sim T^{-\gamma_T}
$, we have
\be
\label{nut}
\nu_T\simeq \left\{
\begin{array}{l}
 \frac \psi2 \left[ 1-\frac{n+2}{n+8}\left(\frac\phi\psi\right)\right]\;,
\qquad \;{\rm if}\quad T \gg T^*
 \\
\frac \psi2\;, \qquad \qquad \qquad \qquad\quad {\rm if}\quad T
\ll T^*,
\end{array} \right.
\end{equation}
with $\gamma_T=2\nu_T$.

This equation shows that, decreasing  the temperature along the
quantum critical trajectory, a crossover temperature $T^*$ exists
which separates two different low-$T$ regimes in the influence
domain of the QCP. The effective correlation length exponent, which
 describes the crossover between these two regimes, can be
easily found rewriting Eq.~(\ref{elstar3}), in terms of the
suitable crossover parameter $\tau=T/T^*$, as
\be
\label{elstarh1} e^{l^*}=[A(n,d)
u_0]^{-\frac12}(T^*)^{-\frac\psi2} h_1(\tau), \ee where the new
scaling function $h_1(\tau)$ is given by
\be
\label{acca1}
h_1(\tau)=\tau^{-\psi/2}[1+\tau^\phi]^{\frac{n+2}{2(n+8)}} \;. \ee
Then,  for the effective exponent $\nu_T^{\rm eff}(\tau)$, we find
\be
\label {nuTeff}
\nu_T^{\rm eff}(\tau)=-\frac{d \ln
h_1(\tau)}{d \ln \tau} = \frac \psi 2 \left[1-\frac{n+2}{n+8}
\frac{\tau^\phi}{1+\tau^\phi}\left(\frac \phi \psi\right) \right]
\ee
which reduces to the asymptotic values in Eq.~(\ref{nut}) for $\tau
\gg1$ and $\tau \ll 1$, respectively.

We now consider the singular part of the free energy density along
the quantum critical trajectory which, using (\ref{elstar3}), can
be written as:
\be
\label{energia}
F_s(r_{0c},T)\sim [A(n,d) u_0]^{\frac d2}
T^{\frac{d\psi}2} \left[1+\left(\frac
T{T^*}\right)^\phi\right]^{\frac d2 \frac{n+2}{n+8}}.
\ee

From this, with $ {C_s(r_{0c},T)}/T \sim T^{-\alpha_T}$, it
immediately follows
\be
\label {alphaT}    \alpha_T\simeq \left\{
\begin{array}{l}
 2 -\frac{d\psi}2 +\frac{n+2}{n+8} \frac{d\phi}2 \quad, \qquad \, {\rm if}\quad T \gg T^*
 \\
 2 -\frac{d\psi}2\quad,  \qquad   \qquad   \qquad  {\rm if}\quad T \ll
 T^*,
\end{array} \right.
\end{equation}
with $\alpha_T<0$ for $d_{cu}^{(q)}<d<4$. Note that  in both cases
the hyperscaling relation $2-\alpha_T=d\nu_T$ is satisfied. As for
the correlation length, an effective specific heat exponent can be
easily obtained from Eq.~(\ref{energia}) as a function of the
crossover parameter $T^*/T$.

Eqs.~(\ref{nut})-(\ref{alphaT}) are particularly interesting
because they show in a transparent way the effects of quantum
critical fluctuations through the shift exponent, which is
strictly related to the Matsubara-frequencies reduction procedure
and hence to the quantum nature of the system under study. Also
the predicted crossover which should occur decreasing the
temperature through $T^*$ is of interest especially because it may
constitute a stimulating suggestion  for experiments.

Notice that we have chosen here to express the asymptotic values
of the exponents $\nu_T$ and $\alpha_T$ in terms of the phase
boundary exponent $\psi$, just with the aim to underline these
important features. The explicit values of $\nu_T$ and $\alpha_T$
to first order in $\eps=4-d$ are presented in Table~\ref{T3}.

As a conclusion of this section, it is worth mentioning that
previous results suggest also that another crossover occurs
on increasing $r_0$ to $r_{0c}$ ($T_c(r_0)\rightarrow 0$ as $r_0
\rightarrow r^-_{0c}$) between the critical regimes found by
approaching the critical line  at fixed $r_0<r_{0c}$ and the QCP
along the quantum critical trajectory ($r_0=r_{0c}, T\rightarrow
0$). It is easy to show that this crossover, like those explored
before, can be described again in terms of effective exponents as
functions of the appropriate crossover parameter $0\leq \tau
'=T_c(r_0)/T\leq 1$, with $T\rightarrow T_c^+(r_0)$ and $r_0
\rightarrow  r_{0c}^-$. This can be performed along the same lines
used for the previous two crossovers, but we prefer to postpone the
problem to the next section where we present  a unified framework
of all crossovers that occur at a fixed $r_0\leq r_{0c}$ on
decreasing the temperature within the disordered phase.

\begin{table}
\caption{Values of $\nu_T$ and $\alpha_T$ to first order in $\eps$ along the quantum critical trajectory.}
\label{T3}
\begin{center}
\begin{tabular}{| c | c | c || c | c |}
\hline
$\vphi(\vk,\omega_l)$ & \multicolumn{2}{c}{$T \ll T^*$} & \multicolumn{2}{c}{$T \gg T^*$}\\
\cline{2-5}
& $\nu_T$ & $\alpha_T$ &  $\nu_T$ & $\alpha_T$ \\
\hline
$\omega_l^2$ & $\ds\frac32-\frac\eps2$ & $-4+\ds\frac72\eps$ &
$\ds\frac{4(n+11)+(n-10)\eps}{4(n+8)}$ & $-\ds\frac{2(n+14)-21\;\eps}{n+8}$\\
$-i\omega_l$ and $|\omega_l|$ & $1-\ds\frac\eps4$ & $-2+2\eps$ &
$\ds\frac{2(n+14)+(n-4)\eps}{2(n+8)}$ &
$-\ds\frac{24-(22-n)\eps}{2(n+8)}$\\
\hline
\end{tabular}
\vspace{0.6cm}
\end{center}
\end{table}

\section{A unified description of crossovers for $r_0\leq r_{0c}$
in terms of  two-parameters effective exponents}

The former analysis shows clearly that the  scenario close to the
QCP, which emerges by approaching the phase boundary along
vertical paths decreasing $T$ at fixed $r_0\leq r_{0c}$ within the
disordered phase, is richer than the one for horizontal
thermodynamic trajectories. Here we want to show that all the
vertical crossovers that take place within the region of the
phase diagram delimited by the critical line and the quantum
critical trajectory, can be globally described  in terms of
two-parameters scaling functions or related effective exponents.
This interesting feature allows one to have a transparent unified
picture of the complex competition between thermal and quantum
fluctuations close to the QCP. Without loss of generality, we
focus on the correlation function (and hence on the directly
related susceptibility) but the crossovers of the other
thermodynamic quantities can be studied similarly. Within this
general framework one can easily reproduce all the asymptotic
behaviors which may have direct experimental interest.

From the basic equation (\ref{elstar}) and the representation
(\ref{5.18}) of the distance from the critical line $t(r_0,T)$, it
is straightforward to check that one  can write
\be
\label{xi2} \xi \simeq \xi_0 [A(n,d) u_0]^{-\frac12}
(T^*)^{-\frac\psi2} H(\tau_1,\tau_2).
\ee
Here
\be
\label{Hcross}
H(\tau_1,\tau_2) =
\tau_2^{-\frac\psi2}(1-\tau_1)^{-\frac12}
[1+\tau_2^\phi(1-\tau_1)^{-\frac\eps2}]^{\frac{n+2}{2(n+8)}}
\ee
is a scaling function of the two natural crossover parameters
(we use here more convenient notations to avoid possible
confusion)
\be
\label{tau}
\tau_1=\frac{T_c(r_0)}{T} \;, \quad \tau_2 =
\frac T{T^*} \;,
\ee
with $0\leq \tau_1\leq1$ and $\tau_2\geq 0$,
where $\tau_1=0$ and $\tau_1=1$ correspond to $r_0=r_{0c}$ and
$r_0<r_{0c}$, respectively.

Since we wish to include in the
analysis also the possibility $T_c(r_0)\rightarrow 0$ as
$r_0\rightarrow r_{0c}^-$, it is now convenient to define the
effective exponent of interest as
\be \label{nueff} \nu_T^{\rm eff}(T,T_c(r_0))=-\frac{d \ln
\xi(T,T_c(r_0))}{d\ln(T-T_c(r_0))} \;. \ee Then, working in terms
of the parameters $\tau_1$ and $\tau_2$, after some tedious but
straightforward calculations, we obtain for $\nu_T^{\rm eff}$ the
noteworthy expression

\be
\label{nueff12}
\nu_T^{\rm eff}(\tau_1,\tau_2)=\frac\psi2
\frac{1-\tau_1}{1-\tau_1^\psi} \left\{
1+\frac{n+2}{2(n+8)}\eps \frac{\tau_2^\phi}
{(1-\tau_1^\psi)^{\eps/2}+\tau_2^\phi} \left[1-
{2(1-\tau_1^\psi)\over{ \psi\eps}} \right]   \right\}.
\ee

 Eqs.~(\ref{Hcross}) and (\ref{nueff12}) are the basic results to
 describe properly all crossovers which occurs approaching
 the critical line along paths parallel to the $T$-axis in the phase diagram.

 We consider explicitly the following asymptotic cases:

(i) $\tau_1\to 1 \qquad (T\to T_c^+(r_0)\neq 0)$.
\\
In this case we easily see that Eq.~(\ref{nueff12}) reduces to
\bea
\label{6.6}
\nu_T^{\rm eff}(\tau_1\to 1,\tau_2)&\simeq& \nu_T^{\rm eff}(\tau_1,\tau_{2c})=\\
\nonumber
 &=&\frac12
\left[1+\frac{n+2}{2(n+8)}\frac\eps{1+\psi^{\eps/2}\tau_{2c}^{-\phi}(1-\tau_1)^{\eps/2}}
\right], \eea where $\tau_{2c}=T_c(r_0)/T^*$. In particular, when
\label {6.7} \be
\delta=\psi^{\eps/2}\tau_{2c}^{-\phi}(1-\tau_1)^{\eps/2}\ll 1 \;,
\ee
the effective exponent (\ref{6.6}) assumes the asymptotic value
\be
\label {6.8}
\nu_T^{\rm eff}(\tau_1\rightarrow 1
,\tau_{2c})\simeq\frac12\left[1+\frac{n+2}{2(n+8)}\eps\right],
\ee
which reproduces the W result in (\ref{5.21}).
 On the contrary, when $\delta \gg 1$,
Eq.~(\ref{6.6}) gives the MF value
\be
\label{nuMF}
\nu_T^{\rm eff}(\tau_1\rightarrow 1,\tau_{2c})\simeq\frac12. \ee
Of course the MF-W crossover line is determined by $\delta\sim1$,
which yields, consistently,  the Ginzburg line found before.

 (ii) $\tau_1=0 \qquad (T_c(r_0)=0,  r_0=r_{0c})$.
\\
In this case Eq.~(\ref{nueff12}) reduces to
\be
\label {nueffettivo1}
\nu_T^{\rm
eff}(\tau_1=0,\tau_{2})=\frac\psi2 \left[1-\frac{n+2}{n+8}
\frac{\tau_2^\phi}{1+\tau_2^\phi}\left(\frac\phi\psi\right)\right] \;,
\ee
 and, in agreement with Eq.~(\ref{nut}), we get
\be
\label {nueffettivo2}  \nu_T^{\rm eff}(\tau_1=0,\tau_{2}\gg 1)
\simeq \frac\psi2
\left[1-\frac{n+2}{n+8}\left(\frac\phi\psi\right)\right]\;,
\ee
 and
\be
\label {nueffettivo3}
 \nu_T^{\rm eff}(\tau_1=0,\tau_{2}\ll 1)\simeq  \frac\psi2 \;.
\ee

(iii) $0\leq\tau_1\leq 1, \qquad \tau_2\simeq \tau_{2c}=
T_c(r_0)/T^* \rightarrow 0$  \quad ($r_0\rightarrow r_{0c}^-$).
\\
Under these conditions, a crossover between the classical W regime
and that along  the quantum critical trajectory  occurs as
$T\rightarrow T_c^+(r_0)$ with $T_c(r_0)\rightarrow 0$ for
$r_0\rightarrow r^-_{0c}$.

From the general equation (\ref{nueff12}), one finds
\begin{eqnarray} \label {nueff1}
\nu_T^{\rm eff}(\tau_1,\tau_{2c}\rightarrow 0 )&=&\frac\psi2
\frac{1-\tau_1}{1-\tau_1^\psi} \left\{
1+\frac{n+2}{2(n+8)}\eps\tau_{2c}^\phi
({1-\tau_1^\psi})^{-\eps/2}
 \left[1-
{2(1-\tau_1^\psi)\over{ \psi\eps}} \right] + O(\tau_{2c}^{2\phi}) \right\}=
 \nonumber \\
&=&\frac\psi2 \frac{1-\tau_1}{1-\tau_1^\psi}+O(\tau_{2c}^{\phi})=
 \left\{
\begin{array}{l}
 {\ds\frac\psi2}\;, \qquad \qquad \tau_1\rightarrow 0
 \\ \\
 {\ds\frac12}\;,  \qquad   \qquad \tau_1\rightarrow 1.
\end{array} \right.
\end{eqnarray}
Since $\psi>1$ for $d_{cu}^{(q)}<d<4$ for all quantum models (see
Table~\ref{T2}), we get
\begin{equation}
\label{range nu}
 \frac12 \leq \nu_T^{\rm
eff}(\tau_1,\tau_{2c}\rightarrow 0 )\leq \frac\psi2\;.
\end{equation}
In particular, for transverse-Ising-like models as $d\rightarrow
3^+$ one has $\nu_T^{\rm eff}(\tau_1,\tau_{2c}\rightarrow 0
)\simeq {1\over {1+\tau_1}}$, so that  ${1\over 2}\leq\nu_T^{\rm
eff}(\tau_1,\tau_{2c}\rightarrow 0 )\leq 1$ and hence
$1\leq\gamma_T^{\rm eff}(\tau_1,\tau_{2c}\rightarrow 0
)\simeq{2\over{1+\tau_1}}\leq 2$. The previous results suggest
that, for systems well described by the model action (\ref
{S1})-(\ref{S3}) with $\vphi (\vec k,\omega_l)=\omega_l^2$,
accurate susceptibility measurements as $T\rightarrow T_c^+(r_0)$
sufficiently close to the QCP should signal an increasing of the
exponent $\gamma_T$ from the value $\gamma_T=1$ to $\gamma_T=2$ as
$T_c(r_0)\rightarrow 0$ when $r_0\rightarrow r_{0c}^-$. This
static ($d\rightarrow 3^+$)-extrapolation prediction appears to be in
very good agreement with available experimental data for quantum
ferroelectrics and other systems with quantum structural phase
transitions \cite{structPT,samara} and for transverse-Ising-type
magnetic materials \cite{erkelens}. It is also worth mentioning
that our static RG results agree with alternative approaches
around $d=3$ based on conventional quantum RG treatments
\cite{sachdev,noi97,noi03} and field-theoretic techniques
\cite{schmeltz}. This constitutes a clear proof of
matching-consistency between static and dynamic theories as
$d\rightarrow 3^+$.

For quantum models with $-i\omega_l$ and $|\omega_l|$,
sufficiently close to the QCP we have $\nu_T^{\rm
eff}(\tau_1,\tau_{2c}\rightarrow 0)\simeq {3\over
4}({1-\tau_1})/({1-\tau_1^{3/2}})$ and hence $ {1\over 2}\leq
\nu_T^{\rm eff}(\tau_1,\tau_{2c}\rightarrow 0 )\leq {3\over 4}$
and  $1\leq\gamma_T^{\rm eff}(\tau_1,\tau_{2c}\rightarrow 0
)\simeq (3/2)({1-\tau_1})/({1-\tau_1^{3/2}})\leq 3/2$.

Coming back to the general equation (\ref {nueff12}) for $0\leq
\tau_1\leq1$ and $\tau_2\geq 0$, with $T\rightarrow T_c^+(r_0)$ we
get
\begin{equation}
\label{range nueff} \nu\leq\nu_T^{\rm eff}(\tau_1,\tau_2)\leq
\psi/2,
\end{equation}
where the correlation length critical exponent
$\nu\equiv\nu_T=\nu_r=\frac
{1}{2}(1+\frac{n+2}{2(n+8)}\eps)$ characterizes the
finite temperature classical W critical regime. With $n=1$ and
$\omega_l^2$-dynamics, as $d\rightarrow 3^+$ we have
\begin{equation}
\label{range gammaeff}
1.17\leq\gamma_T^{\rm
eff}(\tau_1,\tau_2)\leq \psi/2,
\end{equation}
which can be a good starting point for a comparison with
experimental findings \cite{samara,erkelens}. A similar result can
be obtained for other quantum models at $d=3$ by means of an
appropriate use of the general inequality (\ref {range nueff}).


\section{Other crossover lines for $r_0  > r_{0c}$ and the global phase diagram.}

We now focus our attention on the region of the phase diagram to
the right side of the quantum critical trajectory ( $r_0
> r_{0c}$). Here we are sufficiently far from the critical line but
still in the influence domain of the QCP. As mentioned in Sec.
V, also the physics of this regions is fully contained in the
general equation (\ref{elstar}) for the dimensionless correlation
length $\xi/\xi_0\simeq e^{l^* }$ (here we assume $\xi_0=1$). Of
course, to extract the correct physics for $r_0>r_{0c}$ we must
consider the full expression (\ref{tr0T}) for $t(r_0,T)$ which,
for the next developments, can be conveniently written  in terms of
$g=r_0-r_{0c}\ll 1$ (to leading order in $u_0$) as
\be
 \label{fulltr0}
 t(r_0,T)\simeq g + A(n,d)u_0 T^\psi
+ K_d(n + 2)u_0 T^\psi \left[ {{\cal G}\left(
{\frac{g}{{T^{2/\zeta} }}} \right) - {\cal G}(0)} \right], \ee or,
equivalently, as \be \label{fulltr0-2}
 t(r_0,T)\simeq
g + K_d(n+2) u_0 T^{\psi}{\cal G}\left(
{\frac{g}{T^{2/\zeta}}}\right), \ee in view of the definition
(\ref{defs}) of $A(n,d)$.

Due to the peculiar competing effects of the two small parameters
$g$ and $T$ which enter Eqs.~(\ref{fulltr0})-(\ref{fulltr0-2}) for $t(r_0,T)$ and, hence, all the
relevant thermodynamic quantities as susceptibility, specific heat
and so on, different low-temperature regimes may occur. We
consider here the two limit cases $g/ T^{2/\zeta} \ll 1$ and  $g /
T^{2/\zeta}\gg 1$.

(i) $g/ T^{2/\zeta} \ll 1$.

Under this condition, Eq.~(\ref {fulltr0}) yields
\begin{equation}
\label{trot} t(r_0 ,T) \simeq g + A(n,d)u_0 T^\psi \;,
\end{equation}
so that the correlation length is given by
\bea
\label{fullxi}
\xi &\simeq& \left[ {g + A(n,d)u_0 T^\psi  }
\right]^{ - \frac{1}{2}}\times\nonumber
\\ &\times&\left
\{1+(n+8)K_d\left({{u_0}\over{\eps}}\right)T\left[g +A(n,d)u_0
T^\psi\right]^{-\eps /2}\right \}^{\frac{n+2}{2(n+8)}} \;.
\end{eqnarray}

Notice that, when $r_0=r_{0c}$, Eq.~(\ref{fullxi}) reduces to Eq.~(\ref{elstar3})
and hence all the results of
Sec.~\ref{VB}  are reproduced, as expected.

From Eq.~(\ref{fullxi}) two asymptotic regimes appear.

When $g \ll A(n,d)u_0T^\psi$, the properties are essentially
controlled by temperature so that, as $T\rightarrow 0$ with $g>0$,
but $T \gg [A(n,d)u_0]^{-1/\psi}g^{1/\psi}$, one finds for $\xi$
and $C_s/T$ the behaviors already obtained at $r_0=r_{0c}$
involving the crossover temperature $T^*$. This regime will be
called ``renormalized MF ($RMF$) regime'' ($RMF_1$ and $RMF_2$ for
$T\gg T^*$ and $T\ll T^*$, respectively).

In the opposite  case $g \gg A(n,d)u_0T^\psi$, it is easy to check
that
\begin{eqnarray} \label{xi-2}
\xi^{-2}&\simeq&\left[g+A(n,d)u_0T^{\psi}\right]\left\{1+(n+8)K_d\left(
\frac{u_0}{\eps}\right)Tg^{-{\eps/2}}\right\}^{-\frac{n+2}{n+8}}\nonumber
\\  &\simeq& g+A(n,d)u_0T^{\psi},
\end{eqnarray}
and
\be 
\label{Csing} 
\frac{{C_s (r_0 ,T)}}{T}\sim \frac
{d}{2}\psi(\psi-1)A(n,d)u_0\,g^{\frac{d-2}{2}}T^{\psi-2} , 
\ee
where now $g$ dominates and $A(n,d)u_0T^{\psi}$ represents the
leading $T$-dependent deviation from the ($T=0$)-MF behavior
(Q-regime) of the correlation length as $g\rightarrow 0^+$. This
will be called the $Q_1$-regime. Of course, the crossover between
the previous low-$T$ regimes 
 is signaled by the crossover line in the phase diagram
\begin{equation}
\label{Tcroce} T_1(r_0 )=\left[ {A(n,d)u_0 } \right]^{
- \frac{1}{\psi }} \left( {r_0  - r_{0c} } \right)^{ 
\frac{1}{\psi }} \;, \qquad (r_0  \ge r_{0c} ).
\end{equation}
It is worth noting that this is symmetric to the critical line
with respect to the quantum critical trajectory $r_0=r_{0c}$.

(ii) $g \gg  T^{2/\zeta}$.

Now, one needs the leading contribution to ${\cal
G}(g/T^{2\zeta})$ for $g/T^{2\zeta}\gg 1$ in the representation
(\ref{fulltr0-2}). This dependence is different for the three
classes of quantum models here considered
{\cite{sachdev,noi-largeN,sachdev97,noi97}} and hence it is
convenient to discuss separately the three cases.

$(\rm ii)_1\;$ $(\omega_l^2)$-dynamic.

With $g/T^2\gg 1$, it is \cite {sachdev,sachdev97}
\be 
\label{gstorta} 
{\cal G}\left( \frac{g}{T^{2}} \right)\simeq
\Gamma (d/2)
2^{d/2-1}\left(g/T^2\right)^{\frac{d-2}{4}}e^{-\sqrt{g/T^2}} \;.
\ee 
So, for $t(r_0,T)$ we find
\begin{equation}
t(r_0 ,T) \simeq g + \frac{n + 2}{(2\pi)^{d/2}}u_0 T^{d-1} \left
(g/T^2\right )^{\frac{d-2}{4}} e^{-\sqrt{g/T^2}} \;,
\end{equation}
to be compared with Eq.~(\ref{trot}) in the opposite regime.
Then, straightforward calculations show that, as $T\rightarrow 0$,
the correlation length and the singular part of the specific heat
reduce to
\begin{equation}
\label{invcorrel} \xi^{-2}  \simeq  g +  \frac{n +
2}{(2\pi)^{d/2}} u_0 T^{d-1}
\left(\frac{g}{T^2}\right)^{\frac{d-2}{4}}e^{-\sqrt{g/T^2}}\;,
\end{equation}
and
\begin{equation}
\label{sh1} \frac{{C_s (r_0 ,T)}}{T} \sim {d\over {2(2\pi)^{d/2}}}
(n+2)u_0g^{\frac{3d-3}{4}}T^{\frac{d-8}{2}} e^{-\sqrt{g/T^2}} \;.
\end{equation}
Comparing with the corresponding equations (\ref{xi-2}) and
(\ref{Csing}), which are valid within the region $g/T^2\ll 1$ of the phase
diagram below the line $T_1\simeq
[A(n,d)u_0]^{-{1/{(d-1)}}}(r_0-r_{0c})^{1/{(d-1)}}$, we see that,
crossing the additional line
\begin{equation}
T_2 (r_0 ) \simeq(r_0  - r_{0c} )^{\frac{1}{2}} \;,
\end{equation}
a crossover takes place, decreasing $T$, between the $Q_1$-regime to
a new $Q_2$- one characterized by a $T$-dependent deviation from the $(T=0)$-
quantum MF behavior ($Q$-regime) of the correlation length as $g\rightarrow
0^+$ weaker than the simple power law form $A(n,d)u_0T^{d-1}$ which
enters the $Q_1$-regime.

$(\rm ii)_2\;$ $(-i\omega_l)$-dynamics.

Here, with $g/T \gg 1$, we have \cite{noi-largeN,noi97}
\begin{equation}
{\cal G}(g/T) \simeq \frac {1}{2}\Gamma(d/2)e^{-{g/T}} \;,
\end{equation}
so that, for $\xi$ and $C_s (r_0 ,T)/T$, we get
\begin{equation}
\label{invcorrel2} \xi^{-2}  \simeq  g +  \frac{n +
2}{(4\pi)^{d/2}} u_0 T^{d/2} \; e^{-g/T},
\end{equation}
and
\begin{equation}
\label{sh2}
\frac{{C_s (r_0 ,T)}}{T} \sim \frac{d}{2^{d+1}\pi^{d/2}}
(n+2)u_0g^{d/2 +1}T^{\frac{d-8}{4}}e^{-g/T}  \;.
\end{equation}
Then, the line
\begin{equation}
\label{Tcrocecroce}
T_2 (r_0 ) \simeq(r_0  - r_{0c} ) \;,
\end{equation}
signals a crossover between the two quasi-quantum regimes $Q_1$
and $Q_2$ which are characterized by the $T$-dependent deviations
$A(n,d)u_0T^{d/2}$ (Eq.(\ref {xi-2})) and $B(n,d)u_0
T^{\frac{d}{2}}e^{-g/T}$ (Eq.(\ref {invcorrel2})) from the MF
quantum critical behavior of $\xi^{-2}$ as $g\rightarrow 0^+$.

$(\rm ii)_3\;$ $|\omega_l|$-dynamics.

This case, although characterized by the same exponents $\psi$ and
$\zeta$, is sensibly different from the Bose-like one due to the
peculiar effect of the sums over Matsubara frequencies
\cite{sachdev}. Here, with $g/T\gg 1$, one finds indeed
\be \label {gstortabis} {\cal G}\left( \frac{g}{T} \right)\simeq
\frac{\pi}{6}\Gamma (d/2) \zeta(\frac{4-d}{2})\left(\frac{g}{T}
\right)^{-{\frac{4-d}{2}}}  \;. \ee Then, we have
\begin{equation}
\label{invcorrel3}
\xi^{-2}  \simeq  g +  \frac{(n +
2)}{(4\pi)^{d/2}}\frac{\pi}{3} \Gamma (\frac{4-d}{2})u_0 T^{d/2}
(\frac{g}{T})^{-{\frac{4-d}{2}}},
\end{equation}
and
\begin{equation}
\label{sh3}
 \frac{{C_s (r_0 ,T)}}{T} \sim \frac{\pi d}{(4 \pi)^{d/2}}
\Gamma \left(\frac{4-d}{2}\right) (n+2)u_0g^{d-3}.
\end{equation}
Also here, Eq.(\ref{Tcrocecroce}) defines the crossover line from
the $Q_1$-regime  (with $T$-dependent deviation
$A(n,d)u_0T^{d/2}$) in $\xi^{-2}$ to the $Q_2 $-regime
(Eqs.~(\ref{invcorrel3})-(\ref{sh3})) decreasing $T$ to zero at
fixed $g$.

Of course, in all cases, at $T=0$ one has $\xi^{-2}\simeq g$ and
$C_s=0$ ($Q$-regime), as expected.

In summary, for all quantum systems here considered, on the right
of the quantum critical trajectory decreasing the temperature to
zero, one should observe two crossovers among three regimes in the
phase digram, signaled by the  two-lines with equations $T_1(r_0)\simeq [A(n,d)u_0]^{-1/\psi}(r_0-r_{0c})^{1/\psi}$ and
$T_1(r_0)\simeq (r_0-r_{0c})^{\zeta/2}$ (with
$\zeta/2>1/\psi$) ending in the QCP, whose behaviors are
determined by the exponents $\psi$ and $\zeta$ strictly related to
the quantum nature of the original microscopic models. Above the
first line, which is symmetric to the phase boundary within the
region $r_0>r_{0c}$, any system exhibits, essentially, the low-$T$
behavior expected for $r_0=r_{0c}$. Above and below the second
crossover line one finds, for correlation length and
susceptibility, the MF behavior in $r_0-r_{0c}$ (expected at $T=0$
as $r_0\to r_{0c}^+$) but different $T$-dependent corrections
which go to zero more and more rapidly as $T\to 0$. Specifically,
for the inverse susceptibility $\chi^{-1}\sim \xi^{-2}$, crossing
the line $T_2(r_0)$, we find that the $T$-contribution
changes from the power-law shape $A(n,d) u_0 T^\psi$ for all
models, to: $a_1u_0
T^{d-1}\left(\frac{r_0-r_{0c}}{T^2}\right)^{(d-2)/4}
e^{-\sqrt{(r_0-r_{0c})/T^2}}$, if $\vphi(\vk,\omega_l)\simeq
\omega_l^2$ ; $a_2 u_0 T^{d/2} e^{-(r_0-r_{0c})/T}$, if
$\vphi(\vk,\omega_l)\simeq -i\omega_l$; and $a_3 u_0
T^{d/2}[(r_0-r_{0c})/T]^{-(4-d)/2}$, if $\vphi(\vk,\omega_l)\simeq
|\omega_l|$. The constants $a_i (i=1,2,3)$ are defined in
Eqs.~(\ref{invcorrel}), (\ref{invcorrel2}) and (\ref{invcorrel3}),
respectively. Similarly, different regimes occur as $T\to 0$ at
fixed $r_0>r_{0c}$ for the singular part of the specific heat
which goes in any case to zero (in agreement with the Nernst
theorem) with deviation from the Fermi-liquid-like behavior except
for systems with $|\omega_l|$-dynamics (see Eqs.~(\ref{sh1}),
(\ref{sh2}) and (\ref{sh3}), respectively).

\begin{figure}
  \begin{center}
    \includegraphics[width=14cm]{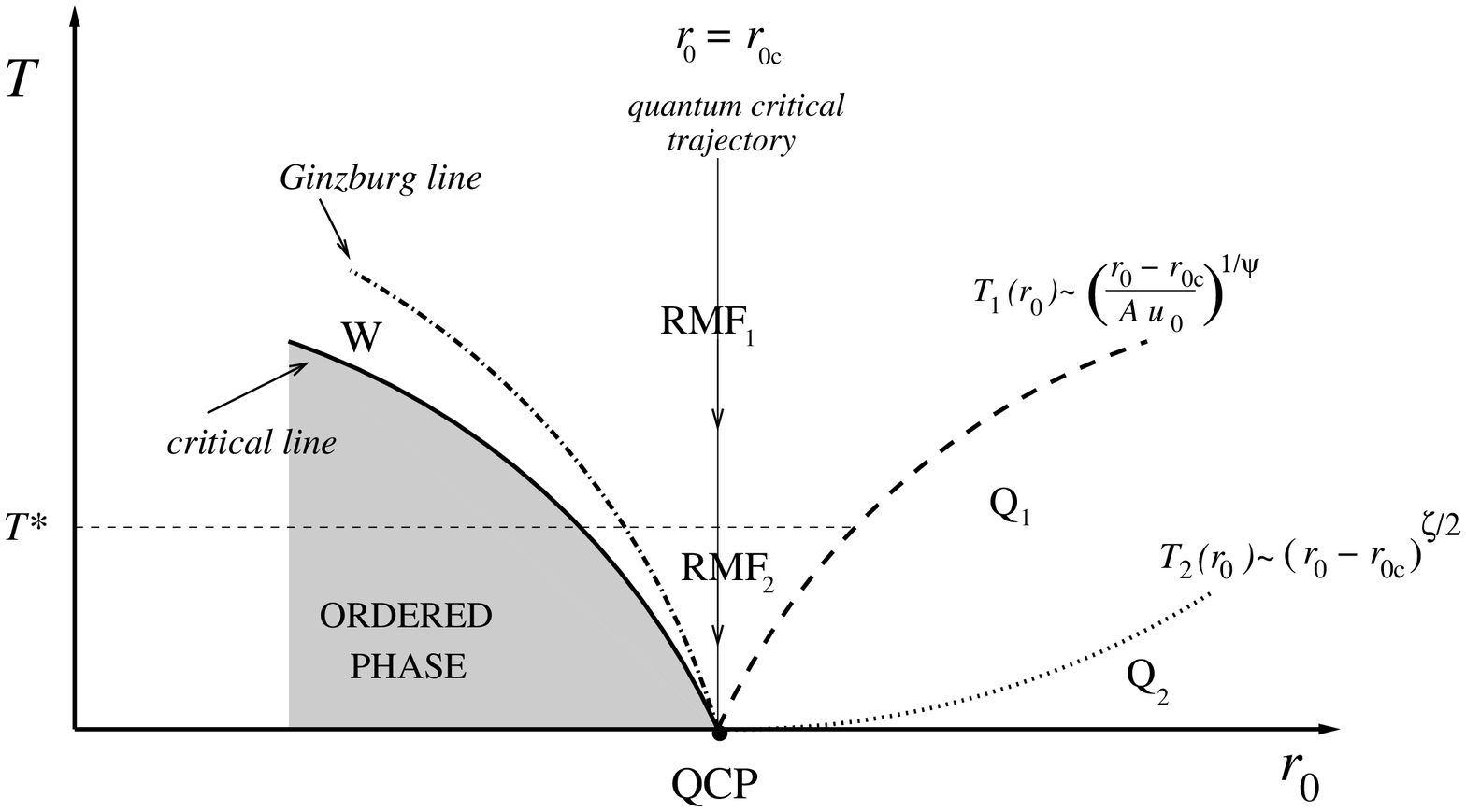}
  \end{center}
  \caption{Qualitative global low-$T$ phase diagram for $(d\lesssim 4)$-dimensional
quantum systems which emerges from our static framework. Here $T$ is the temperature and
$r_0$ the nonthermal control parameter. The continuous line denotes the phase boundary
and the noncontinuous ones indicate crossovers. The shaded region represents the ordered
phase. Whitin the W region, a classical critical behavior takes place (approaching the critical
line along vertical and horizontal paths). The central region is characterized by a crossover between two
different regimes (${\rm RMF}_1$ and ${\rm RMF}_2$) which occurs decreasing $T$
along the quantum critical trajectory, with MF-like
exponents renormalized through the shift exponent $\psi$. This is signaled by the thin dashed horizontal
line $T=T^*$. In the region $Q_1$, the $T$-dependent contributions to the leading MF behavior
in $(r_0-r_{0c})$ of relevant macroscopic quantities has a power law form again related to $\psi$.
The region $Q_2$ corresponds to the disordered quantum regime where the thermal fluctuations
are negligible. The exponent $\zeta$, which marks the crossover between the quasi-quantum
regimes $Q_1$ and $Q_2$ is identified as the dynamical critical exponent $z$.}
  \label{Fig2}
\end{figure}

The qualitative global low-temperature  phase diagram  in the
$(r_0,T)$-plane for quantum models here considered  for
$d_{cu}^{(q)}<d \lesssim 4$, which emerges from our previous
static analysis, is shown in Fig.~\ref{Fig2}.

It is remarkable that the quantum critical scenario, emergent from the
effective static RG treatment here performed, is quite similar
to that obtained in the literature \cite{sachdev,qpts,sondhi,DCqpt} using RG
approaches which involve directly the Matsubara-time axis.

Another key-point to be clarified is the underlying role played by
the dynamical critical exponent $z$ which characterizes the
intrinsic dynamics of a quantum system which exhibits a QPT. In
this connection, since our theory is strictly static in nature,
one may think that no direct information about the different
quantum models is  possible. This is not the case and information
about $z$ can be simply and consistently extracted from our static
results using the general feature in the theory of critical
phenomena that some scaling relations exist which relate also
static and dynamic exponents. Bearing this in mind, by inspection
of the results collected in Table~\ref{T1}, it is evident that the
shift exponent $\psi$ and the exponent $\zeta$, which enter in all
our predictions as a manifestation of the underlying quantum
degrees of freedom, are not independent but are related by 
\be 
\label{zeta} 
\zeta=\frac{d-2}{\psi-1}, 
\ee
 which is valid for
any quantum model considered through this paper. On the other
hand, in conventional quantum RG approaches
\cite{sachdev,qpts,sondhi,DCqpt}, when the dynamical exponent $z$ is
known, for $d+z>4$ the shift exponent $\psi$ is determined in
terms of $z$ through the scaling relation \cite{qpts,millis93}
\be 
\label{psi} 
\psi=\frac{d+z-2}{z} \;. 
\ee 
Eqs.~(\ref{zeta}) and (\ref{psi}), together with  a careful comparison of the global
static phase diagram in Fig.~\ref{Fig2} with the corresponding
ones derived by means of dynamic theories
\cite{sachdev,qpts,millis93}, allow us to identify the exponent
$\zeta$ as  the appropriate dynamic exponent $z$ forthe  quantum
systems under study.

This identification establishes a bridge between our unified
static analytical predictions and the conventional dynamic
scenario. For instance, the crossover line $T_2(r_0)$ in
Fig.~\ref{Fig2} can be also obtained from our leading order solution
$\xi \sim (r_0-r_{0c})^{-1/2}$ setting (with $\zeta=z$) $\xi
T^z\sim 1$, which coincides with that found in literature
\cite{sachdev,qpts,sondhi,DCqpt,vojta-rev,GSI,millis93}, signaling the so-called
quantum $(T\xi^z \ll 1)$-to-classical $(T\xi^z \gg 1)$ crossover.
Then, when $T\ll T_2(r_0)$, we are in the ``disordered
quantum regime''  where the physics is essentially quantum in
nature in the sense that the fluctuations on scale $\xi$ have
energies much greater than $K_B T$ (where $K_B$ is the Boltzmann
constant here assumed equal to unity).

Within this scenario, the peculiar region around the quantum
critical trajectory in the phase diagram, above the crossover line
$T_1(r_0)$, defines the usual ``quantum critical
region'' characterized by vertical path classical exponents
renormalized as a consequence of the QCP influence.

\section{Conclusions}

In summary, in this paper we have derived, within a general
non-conventional framework, the low-$T$ quantum critical scenario
and the crossovers induced by the interplay of thermal and quantum
critical fluctuations for three wide classes of systems which
exhibit a QCP, also named a ``black hole'' in the phase diagram
\cite{modernphys}. This has been performed, close to and below the
classical upper critical dimensionality, by using an effective
static treatment which combines a preliminary integration over
degrees of freedom with non-zero Matsubara frequencies, to obtain
a classical GL free energy functional with effective $T$-dependent
coupling parameters, and then the genuine Wilson RG philosophy.
This allowed us to extract the quantum critical properties,
crossovers and the global phase diagram of the original quantum
systems. In our intrinsically static RG picture, the explicit
dependence of the effective coupling parameters on temperature and
the original microscopic ones, played a crucial role as a result
of the competition between the classical and quantum critical
worlds. The emergent phase diagram was found to display all the
relevant features currently obtained via more familiar approaches.
Nevertheless, additional low-temperature crossovers were found as
a further manifestation of the QCP influence. It is also worth
emphasizing another relevant result of our static approach. As
well known, the key feature which distinguishes the quantum and
the most familiar  finite temperature phase transitions is that,
while the intrinsic dynamics of a quantum system is irrelevant for
the latter, it plays a crucial role in the former. The link
between statics and dynamics close to a continuous QPT is usually
measured by the value of the dynamical critical exponent $z$, that
describes the relative scaling of the time and the length scales
in the problem. Thus, settling the value of $z$ is of a great
interest, especially to distinguish different quantum universality
classes. This objective is traditionally achieved using dynamic
theories \cite{sachdev,qpts,sondhi,DCqpt,vojta-rev,GSI}. However, although
intrinsically static, our RG analysis allowed us to obtain
information about $z$ through its identification with a new
exponent $\zeta$ which arises from the degrees of freedom
reduction procedure as strictly related to the shift exponent
$\psi$ which characterizes the low-$T$ shape of the phase
boundary.

In conclusion, we hope that our simple approach may give further
insight into the topical subject of QPTs and the effects of
competition between thermal and quantum critical fluctuations
moving in the phase diagram along appropriate thermodynamic
trajectories to approach QCPs. 
On this matter, we believe that a relevant feature of our approach is 
that it allows us to work within a single parameter space in contrast with 
the usual one where the temperature enters explicitly the RG flows. As it is 
well known \cite{sachdev,qpts,sondhi,DCqpt,vojta-rev,GSI,noi97,millis93,noi03,noi05},
to extract complete  physical information, 
the renormalization of the temperature forces to perform  the rather unnatural 
change of the renormalized original coupling parameter $u(l)$ (see Eq.~(\ref{S3}))
in the new one $v(l)=u(l)T(l)$ when the rescaling parameter $l$ is sensibly 
increased by iteration of the RG transformation. It is just this feature 
that implies inevitably the traditional step by step procedure and prevent, 
in our opinion, a unified and controllable description of the crossovers in 
the influence domain of a QCP. More serious problems on physical grounds 
emerge in the conventional RG treatments when quenched disorder is present 
\cite{sachdev,revDC}.  We think that the key idea of our method may be usefully employed, 
especially in this more complex situation, for properly exploring
quenched disorder effects on quantum criticality by overcoming the
well known troubles \cite{sachdev,revDC} related to the
Matsubara-time direction introduced by path-integral
representation as  an expression of the non-commutability of the
operators which enter the microscopic Hamiltonian.

It is also worth mentioning that the present scenario close to the QCP, 
as the previous ones in literature \cite{sachdev,qpts,sondhi,DCqpt,vojta-rev,GSI,millis93},
is strictly valid to the one-loop approximation. Of course, corrections 
to the Fisher exponent $\eta$, and hence to the related ones via the usual scaling 
relations, are expected to higher order approximations. However, 
we believe that the previous physical scenario will remain qualitatively 
unchanged for all quantum systems of interest.


\begin{thebibliography}{99}
\bibitem{footnote} The index $r$ in the
exponents signals that the critical line is approached by
variation of the non-thermal control parameter $r_0$ at fixed $T$.
A different index ($T$) will be used when vertical
paths are used, i.e. trajectories parallel to the $T$-axis in the phase diagram.
\end{thebibliography}
\end{document}